\documentclass[preprint,prd,amsmath,amssymb,amsthm,nofootinbib]{revtex4}
\usepackage{xcolor}
\usepackage[mathscr]{euscript}
\usepackage{bm}
\usepackage{graphicx}
\usepackage[caption=false]{subfig}
\usepackage[colorlinks=true,linkcolor=red]{hyperref}
\usepackage{ulem}

\newcommand{\bea}{\begin{eqnarray}}
\newcommand{\eea}{\end{eqnarray}}
\newcommand{\beq}{\begin{equation}}
\newcommand{\eeq}{\end{equation}}

\begin{document}

\title{The Lamb shift in the BTZ spacetime}
\author{Yaqian Yu$^{1}$~\footnote{1371254089@qq.com}, Jialin Zhang$^{1,2}$~\footnote{Corresponding author. jialinzhang@hunnu.edu.cn}and Hongwei Yu$^{1,2}$~\footnote{Corresponding author. hwyu@hunnu.edu.cn}}
\affiliation{
$^1$ Department of Physics and Synergetic Innovation Center for Quantum Effects and Applications, Hunan Normal University, 36 Lushan Rd., Changsha, Hunan 410081, China\\
$^2$ Institute of Interdisciplinary Studies, Hunan Normal University, 36 Lushan Rd., Changsha, Hunan 410081, China}

\begin{abstract}
We study the Lamb shift of a two-level atom arising from its
coupling to the conformal massless scalar field, which satisfies the Dirichlet boundary condition, in the Hartle-Hawking vacuum in the BTZ spacetime, and find that the Lamb shift in the BTZ spacetime  is structurally   similar to that of a uniformly accelerated atom near a perfectly reflecting boundary in (2+1)-dimensional flat spacetime. Our results show that the Lamb shift is suppressed  in the BTZ spacetime as compared to that  in the flat spacetime as long as the transition wavelength of the atom is much larger than $AdS$ radius of the BTZ spacetime while it can be either suppressed or enhanced if  the transition wavelength of the atom is much less than $AdS$ radius, depending on the location of the atom. In contrast, the Lamb shift is always suppressed very close to the horizon of the BTZ spacetime and remarkably it reduces to that in the flat spacetime as the horizon is approached although the local temperature blows up there.
\end{abstract}

\maketitle

\section{Introduction}
The Lamb shift, which describes a subtle energy level shift of an atom, was first discovered in experiment in 1947~\cite{Lamb:1947} and later theoretically explained as arising from the coupling of the atom with fluctuating quantum fields in vacuum. The Lamb shift is  regarded as  one of the most remarkable effects predicted in quantum theory and marks the beginning of modern quantum electrodynamics~\cite{Dirac:1989}.
So far, the Lamb shift has been investigated  in various circumstances, such as in the presence of cavities~\cite{Meschede:1990}, in a thermal bath~\cite{Barton:1972,Farley:1981,Zhu:2009},
in de Sitter (dS) and the Schwarzschild black hole spacetimes~\cite{Zhou:2010-1,Zhou:2010-2}, as well as for atoms in noninertial motion~\cite{Audretsch:1995,Passante:1998,Rizzuto:2007,Zhu:2010}.   These studies show that the Lamb shift is singularly impacted by the topology and structure of spacetime,  the motion status of the atom and the ambient thermal radiation.

In this paper, we are interested in the Lamb shift in the BTZ spacetime, which is an exact  solution of the Einstein equation in (2+1)-dimensional gravity  found by Ba\~{n}ados, Teitelboim, and Zanelli (BTZ) in 1992~\cite{BTZ-1}.  The BTZ solution has attracted a lot of attention since its discovery, as it is generally believed  that the general relativity in
(2+1) dimensions can be considered as a quite useful laboratory for
exploring the foundations of classical and quantum gravity after the seminal work of Deser et al.~\cite{Deser:1984,Deser:1988}.  It has been found that the BTZ solution displays  interesting features  different from  black holes in other dimensions, such as the absence of a curvature singularity at the origin and the lack of global hyperbolicity~\cite{Lifschytz:1994,Carlip:1995}, as well as  clear advantages  such as an explicit expression for the Green's function of quantum fields and the simplicity of some exact analytical calculations~\cite{Lifschytz:1994,Binosi:1999}. Interesting quantum phenomena associated with the BTZ spacetime,  for instance, the response  of the Unruh-DeWitt (UDW) particle detectors~\cite{Hodgkinson:2012},  quantum fluctuations~\cite{ Pourhassan:2017qxi}, entanglement harvesting~\cite{Zhjl:2018}, anti-Unruh effect~\cite{Zhjl:2020,DeSouzaCampos:2020ddx,Robbins:2021ion} and holographic complexity~\cite{Emparan:2021hyr},
have been explored.  As a further step, we plan to investigate, in the present paper, the Lamb shift for a two-level atom arising from its coupling with the fluctuating conformal  massless scalar fields in vacuum in the BTZ spacetime, hoping to further understand the properties of the BTZ spacetime in terms of the Lamb shift.

Our calculation of the Lamb shift will be  carried out  with the elegant formalism proposed by Dalibard, Dupont-Roc, and Cohen-Tannoudji (DDC)~\cite{DDC-1,DDC-2}, which  allows for a separation of  contributions
of vacuum fluctuations and radiation reaction to an atomic observable  by adopting  a symmetric operator ordering between the operators of the atom and the field.
The  paper is organized as follows. We begin in Sec.~{\ref{sec2}}  by  presenting the basic formulae for the relative radiative energy shift of a two-level atom following the DDC approach. In Sec.~{\ref{sec3}}, we compute with the DDC approach the Lamb shift  for a uniformly accelerated atom in the (2+1)-dimensional flat spacetime with  a perfectly reflecting boundary.  We  then calculate the Lamb shift for the atom in coupling with conformal massless scalar fields in the Hartle-Hawking vacuum in the BTZ spacetime in Sec.~{\ref{sec4}}, and compare it  with that of a uniformly accelerated atom near a perfectly reflecting boundary.  The properties of the Lamb shift are analyzed  not only by the analytical approximations in some special cases but also by  numerical computation. Finally, we end with conclusions in Sec.~{\ref{sec5}}.

For convenience,   the natural units $\hbar=c=k_B=8G=1$ and the
metric signature $(-1,1,1)$ are adopted throughout this paper.

\section{The basic formalism}
\label{sec2}
Let us now consider a two-level atom locally interacting with a fluctuating  conformal
massless scalar field $\phi (x( \tau ))$ in vacuum. For simplicity, the worldline of the atom is
denoted by $x({\tau})$ which is parameterized by its proper time $ {\tau}$.
In Dicke's notation, the Hamiltonian of the atom  $H_{A}(\tau)$
 can be
written  as
\begin{equation}\label{pf1}
H_{A}(\tau)=\omega_{0}R_{3}(\tau)\;,
\end{equation}
while that of the scalar field  $H_{F}(\tau)$  as
\begin{equation}
H_{F}(\tau)=\int d^{3}k~\omega_{\bf{k}} a_{\bf{k}}^{\dagger} a_{\bf{k}} \frac{dt}{d\tau}\;,
\end{equation}
where $\omega_0$ denotes the energy gap between the ground state $|-\rangle$ and excited state $|+\rangle$ of the atom, $R_3(0)=({1}/{2})(|+\rangle\langle+|-|-\rangle\langle-|)$~\footnote{ The time evolution of the  atomic operators, which will be given later, can be obtained from the Heisenberg equations.},  and $a^{\dagger}_{\bf{k}},a_{\bf{k}}$ are the creation and annihilation operators of the scalar field.  The interaction Hamiltonian for the atom-field coupling is assumed to be~\cite{Audretsch:1995}
\begin{equation}
 H_I(\tau )= \mu R_2( \tau ) \phi (x( \tau ))\;,
\end{equation}
where $\mu$ is a small coupling constant and $R_2(0)=(i/2)[R_{-}(0)-R_{+}(0)]$ with $R_{-}(0)=|-\rangle\langle+|,~R_{+}(0)=|+\rangle\langle-|$.

To obtain the energy  level shifts of the two-level atom, we begin  with the Heisenberg equations of motion of  dynamical variables of the atom and the field. For an arbitrary atomic observable $O(\tau)$, the Heisenberg equation  of motion is given by
 \begin{equation}\label{Hei-0}
{{d\over{d \tau }}O( \tau )}=-i[O( \tau ),H_{A}(\tau)+H_{I}(\tau)]\;.
\end{equation}
In the  solution of the equation of motion,  we can split the atom and field operators into two parts~\cite{Audretsch:1995}: the free part that exists even when there is no coupling between the atom and the field and the source part that is induced by the interaction and characterized by the coupling constant, that is
\begin{equation}
R_3(\tau)=R_3^f(\tau)+R_3^s(\tau),~R_{\pm}(\tau)=R_{\pm}^{f}(\tau)+R_{\pm}^{s}(\tau),~\phi(x(\tau))=\phi^{f}(x(\tau))+\phi^{s}(x(\tau))\;,
\end{equation}
where superscripts $f$ and $s$ denote the free and  source part  respectively. A integration of the Heisenberg equations yields~\cite{Audretsch:1994}
\begin{align}
&R_3^f(\tau)=R_3^f(0),~R_3^s(\tau)=i\mu\int_0^\tau {d\tau'}\phi^f(x(\tau'))[R_2^f(\tau'),R_3^f(\tau)]\;,\nonumber\\
&R_\pm^f(\tau)=R_\pm^f(0)e^{\pm{i}\omega_0\tau},~R_\pm^s(\tau)=i\mu\int_0^\tau {d\tau'}\phi^f(x(\tau'))[R_2^f(\tau'),R_\pm^f(\tau)]
\end{align}
and
\begin{align}
&a_{\bf{k}}^f(t(\tau))=a_{\bf{k}}^f(t(0))e^{-i\omega_{\bf{k}}[t(\tau)-t(0)]},~a_{\bf{k}}^s(t(\tau))=i\mu\int_0^\tau {d\tau'}R^f_2(\tau')[\phi^f(x(\tau')),a_{\bf{k}}^f(t(\tau))]\;.
\end{align}
 Here, $R_2(\tau)=R_2^f(\tau)+R_2^s(\tau)=(i/2)[R_{-}^f(\tau)-R_{+}^f(\tau)+R_{-}^s(\tau)-R_{+}^s(\tau)]$ leads to  $R_2^f(\tau)=(i/2)[e^{-i\omega_0\tau}R_{-}(0)-e^{i\omega_0\tau}R_+(0)]$ and $R_2(0)=R_2^f(0)$ as expected. According to  the formalism of DDC~\cite{DDC-1,DDC-2}, one can identify the contribution of  vacuum  fluctuations  (which is related to the free part of the field and denoted by subscript ``$vf$") and  that of radiation
reaction (which is related to the source part of the field and denoted by subscript ``$rr$") to the rate of change of $O(\tau)$ by adopting a symmetric ordering between the atom and the field variables, i.e., the equation of motion~(\ref{Hei-0}) can be recast as
\begin{equation}
\frac{dO(\tau)}{d\tau}=\Big(\frac{dO(\tau)}{d\tau}\Big)_{vf}+\Big(\frac{dO(\tau)}{d\tau}\Big)_{rr}\;,
\end{equation}
with
\begin{equation}\label{evf-1}
\Big(\frac{dO(\tau)}{d\tau}\Big)_{vf}=\frac{1}{2}i\mu\Big(\phi^{f}(x(\tau))\big[R_{2}(\tau),O(\tau)\big]+\big[R_{2}(\tau), O(\tau)\big]\phi^{f}(x(\tau))\Big)\;,
\end{equation}
and
\begin{equation}\label{err-1}
\Big(\frac{dO(\tau)}{d\tau}\Big)_{rr}=\frac{1}{2}i\mu\Big(\phi^{s}(x(\tau))\big[R_{2}(\tau), O(\tau)\big]+\big[R_{2}(\tau),O(\tau)\big] \phi^{s}(x(\tau))\Big)\;.
\end{equation}
Taking the average value of Eqs.~(\ref{evf-1}) and~(\ref{err-1}) over the vacuum state of the scalar field, one obtains~\cite{Audretsch:1995}
\begin{equation}
\Big\langle0\Big|\Big(\frac{dO(\tau)}{d\tau}\Big)_{{vf},{rr}}\Big|0\Big\rangle=i\big[H^{eff}_{{vf},{rr}}(\tau),O(\tau)\big]+\mathrm{non\mbox{-}Hamiltonian\;terms}\;,
\end{equation}
where the effective Hamiltonian to the order $\mu^2$ reads
\begin{equation}\label{Heffvf-1}
H^{eff}_{vf}( \tau) ={1\over2}i \mu ^2\int^ \tau _{ \tau _0}d \tau'C^F(x( \tau ),x( \tau'))[R_2^f( \tau'),R_2^f( \tau)]\;,
\end{equation}
\begin{equation}\label{Heffrr-1}
 H^{eff}_{rr}(\tau)=-{1\over2}i \mu ^2\int^ \tau_{ \tau_0}d \tau' \chi ^F(x(\tau),x(\tau'))\{R_2^f(\tau'),R_2^f(\tau)\}\;.
 \end{equation}
Here, $C^F(x( \tau ),x( \tau '))$ and $\chi ^F(x( \tau ),x( \tau '))$  are respectively the symmetric correlation and linear susceptibility function of the field, which are defined as
\begin{equation}\label{cf-1}
C^F(x( \tau ),x( \tau') ):={1\over2}\langle0|\{ \phi ^f(x( \tau )),
\phi ^f(x( \tau'))\}|0\rangle\;,
 \end{equation}
\begin{equation}\label{chi-1}
 \chi ^F(x( \tau ),x( \tau')): ={1\over2}\langle0|[ \phi ^f(x( \tau )), \phi ^f(x( \tau'))]|0\rangle\;.
\end{equation}
Then averaging Eq.~(\ref{Heffvf-1}) and Eq.~(\ref{Heffrr-1}) over the atomic state $|b\rangle$ (here $b\in\{-,+\}$), we obtain the contributions of vacuum fluctuations and radiation reaction to the energy shift of level $|b\rangle$~\cite{Audretsch:1995}
\begin{equation}\label{ebvf}
  (\delta E_b)_{vf}=-i \mu ^2\int^ \tau _{ \tau _0}d \tau 'C^F(x( \tau),x( \tau')) \chi _b^A( \tau , \tau ')\;,
\end{equation}
\begin{equation}\label{ebrr}
   (\delta E_b)_{rr}=-i \mu ^2\int^ \tau _{ \tau _0}d \tau ' \chi ^F(x( \tau),x( \tau')) C_b^A( \tau , \tau ')\;,
\end{equation}
where the symmetric correlation function of the atom is
\begin{align}
 C_b^A( \tau , \tau'):=\frac{1}{2}\langle b|\{R_{2}^f(\tau),R_{2}^f(\tau')\}| b\rangle=\frac{1}{2} \sum_{d}|\langle b|R_{2}(0)| d\rangle|^{2}\big[e^{i \omega_{b d}(\tau-\tau^{\prime})}+e^{-i \omega_{b d}(\tau-\tau^{\prime})}\big]\;,
\end{align}
and the atomic linear susceptibility
\begin{align}
\chi_b^A(\tau, \tau'):=\frac{1}{2}\langle b|[R_{2}^f(\tau),R_{2}^f(\tau')]| b\rangle=\frac{1}{2} \sum_{d}|\langle b|R_{2}(0)| d\rangle|^{2}\big[e^{i \omega_{b d}(\tau-\tau^{\prime})}-e^{-i \omega_{b d}(\tau-\tau^{\prime})}\big]\;.
\end{align}
Note that $\omega_{bd}=\omega_{b}-\omega_{d}$ is the atomic energy  gap between the two levels.  In particular, $\omega_{0}=|\omega_{b}-\omega_{d}|$ when $b\neq{d}$, and the sum over $d$ should extend over a complete set of atomic states.
For a two-level atom, the Lamb shift as the relative  energy shift is given by
\begin{equation}\label{delta-all0}
\Delta=(\delta E_+)_{vf}+(\delta E_+)_{rr}-(\delta E_-)_{vf}-(\delta E_-)_{rr}\;.
  \end{equation}
 In what follows, we will estimate the Lamb shift of the atom with the statistical functions~(\ref{cf-1}) and~(\ref{chi-1}) evaluated along its world line.

\section{The Lamb shift for an accelerated atom near a reflecting
boundary}
\label{sec3}

\begin{figure}[!htbp]
\centering
\includegraphics[width=0.6\textwidth]{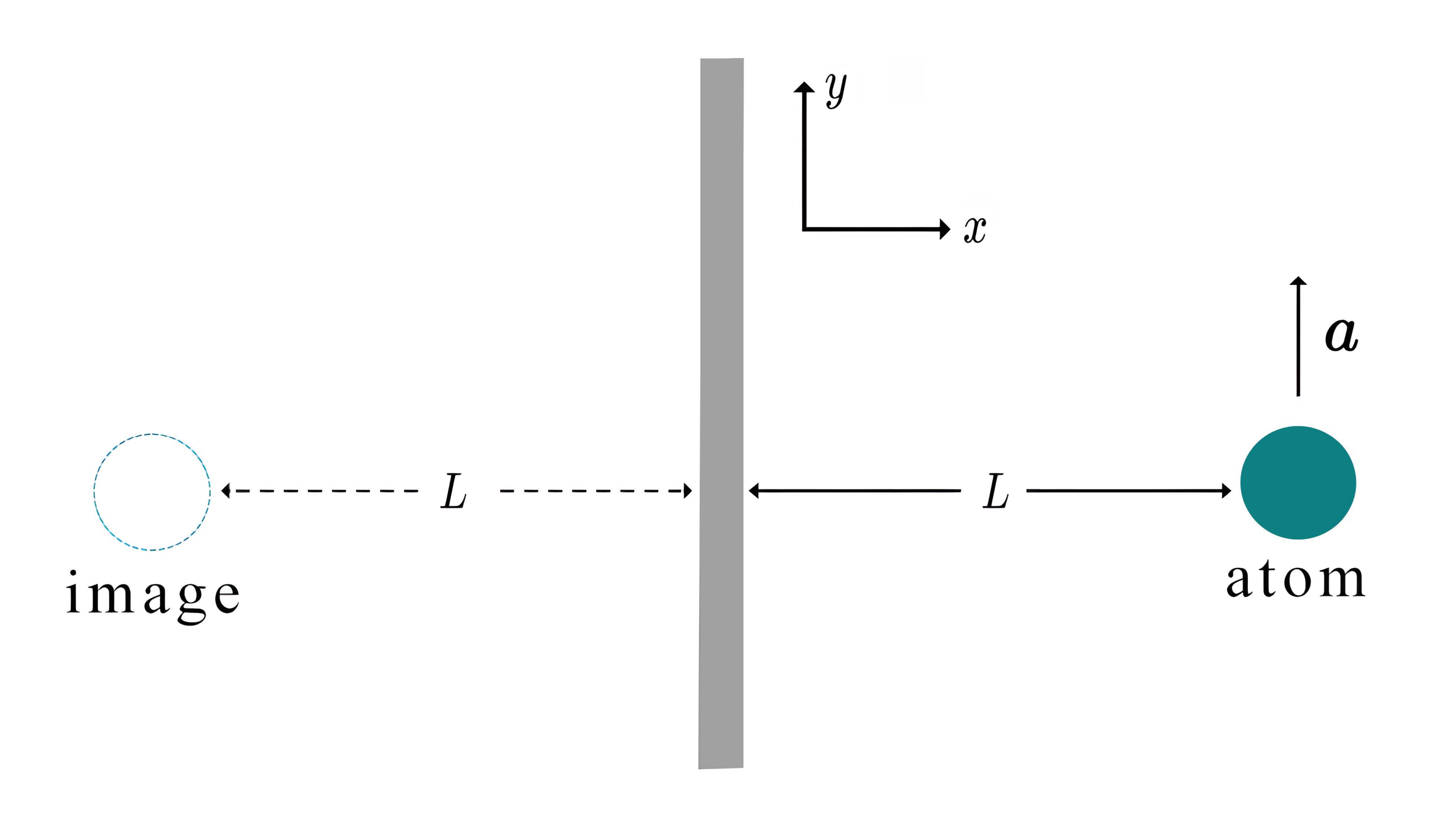}
\caption{An atom is uniformly accelerated along the $y$-axis  with a proper distance $L$ from a reflecting boundary  in (2+1)-dimensional flat spacetime.}\label{BD}
\end{figure}

Let us first consider the Lamb shift in a simpler situation for a later comparison with that in the BTZ spacetime, i.e., that of a uniformly accelerated atom near a perfectly reflecting boundary in the (2+1)-dimensional flat spacetime.
Suppose that the reflecting boundary locates at $x=0$,  and the spacetime trajectory  of the accelerated  two-level atom parameterized by its proper time $\tau$ is given by (see Fig.~(\ref{BD}))
\begin{equation}\label{trajectory}
t(\tau)=\frac{1}{a} \sinh (a \tau), \quad x(\tau)=L,\quad y(\tau)=\frac{1}{a} \cosh (a \tau),
\end{equation}
where $a$ denotes the constant proper acceleration
along the $y$ axis and $L$ is the distance between the atom and the reflecting boundary.

By the method of images~\cite{Birrell:1984}, one can obtain the Wightman function for vacuum massless scalar fields in (2+1)-dimensional Minkowski spacetime with the presence of the reflecting boundary
\begin{align}\label{Green-pb}
	G^{+}_{BD}(x(\tau),x(\tau'))=&\langle0|\phi ^f(x)\phi ^f(x')|0\rangle
	\nonumber\\=&\frac{1}{4 \pi} \frac{1}{\sqrt{\big(x-x^{\prime}\big)^{2}+\big(y-y^{\prime}\big)^{2}-\big(t-t^{\prime}-i \epsilon\big)^{2}}}
	\nonumber\\&-\frac{1}{4 \pi} \frac{1}{\sqrt{\big(x+x^{\prime}\big)^{2}+\big(y-y^{\prime}\big)^{2}-\big(t-t^{\prime}-i \epsilon\big)^{2}}}\;.
\end{align}
Substituting the spacetime trajectory~(\ref{trajectory}) of the two-level atom into Eq.~(\ref{Green-pb}), we have
\begin{align}\label{Greenpbac}
	G^{+}_{BD}(x(\tau),x(\tau'))=\frac{a}{8 \pi} \frac{1}{\sqrt{-\big[\sinh\big(\frac{a \Delta\tau}{2}-i\epsilon \big) \big]^{2}}}-\frac{a}{8 \pi} \frac{1}{\sqrt{-\big[\sinh \big(\frac{a \Delta\tau }{2}-i\epsilon \big) \big]^{2}+(aL)^{2}}}\;
\end{align}
with $\Delta\tau=\tau-\tau'$.

Now we will calculate the corrections to the Lamb shift because of the presence of the boundary and acceleration. That is, what we  actually calculate is the Lamb shift relative to that when the atom is at rest  in a free space.  In (3+1)-dimensional spacetime, directly calculating the relative Lamb shift instead of the total one  avoids the tricky issue of regularizing the Lamb shift in free space. Normally, in a non-relativistic quantum field theoretic  approach such as what we are using here, a suitable cutoff is  needed in the regularization. However, let us note that  no regularization is actually required  for the (2+1)-dimensional case, as now  the Lamb shift in free space is finite. Nevertheless,  since our main concern in the present paper is the corrections due to the boundary and acceleration here and the BTZ spacetime later, we only examine the relative Lamb shift.

So, in our calculation of the atomic energy-level shifts, we shall use the  Wightman function  $G^{+R}_{BD}(x(\tau),x(\tau'))$ which is obtained  by subtracting  from the Wightman function~(\ref{Greenpbac}) that of the massless scalar field  in the  Minkowski vacuum in (2+1)-dimensional flat spacetime given by~\cite{Takagi:1986}
\begin{equation}\label{Ffla}
G^{+}_{FLA}(x(\tau),x(\tau'))=\frac{1}{4\pi\sqrt{(x-x')^2+(y-y')^2-(t-t'-i \varepsilon )^2}}=\frac{1}{4\pi\sqrt{-(\Delta\tau-i \varepsilon )^2}}\;
\end{equation}
to give
\begin{equation}
G^{+R}_{BD}(x(\tau),x(\tau'))=G^{+}_{BD}(x(\tau),x(\tau'))- G^{+}_{FLA}(x(\tau),x(\tau'))\;.
\end{equation}
Then, two statistical functions of the field can be written as the  symmetrized and anti-symmetrized  Wightman functions:
\begin{equation}
C^{F}_{BD}(x(\tau),x(\tau'))=\frac{1}{2}\big[G^{+R}_{BD}(x(\tau),x(\tau'))+G^{+R}_{BD}(x(\tau'), x(\tau))\big]\;,
\end{equation}
\begin{equation}
\chi^{F}_{BD}(x(\tau),x(\tau'))=\frac{1}{2}\big[G^{+R}_{BD}(x(\tau),x(\tau'))-G^{+R}_{BD}(x(\tau'), x(\tau))\big]\;.
\end{equation}
Substituting the above statistical functions into Eqs.~(\ref{ebvf}) and (\ref{ebrr}),
we can separately calculate the contributions of vacuum fluctuations and radiation reaction to the energy shift of level $b$.
Generally, it is  difficult to  directly compute the integrals involved  by using the residual theorem and the contour integration technique. So, we now first write the statistical functions as the following  Fourier integrals
\begin{align}\label{cf-11}
C^{F}_{BD}(x(\tau),x(\tau'))=\frac{1}{4\pi}\int_{-\infty}^{\infty}\big[e^{i \omega(\tau-\tau')}F^{R}_{BD}(\omega)+e^{-i \omega(\tau-\tau')}F^{R}_{BD}(\omega)\big]d\omega\;,
\end{align}
\begin{align}\label{chi-11}
\chi^{F}_{BD}(x(\tau),x(\tau'))=\frac{1}{4\pi}\int_{-\infty}^{\infty}\big[e^{i \omega(\tau-\tau')}F^{R}_{BD}(\omega) -e^{-i \omega(\tau-\tau')}F^{R}_{BD}(\omega)\big]d\omega\;,
\end{align}
where $F^{R}_{BD}(\omega)$ represents the Fourier transform of the Wightman function $G_{BD}^{+R}(\Delta\tau)$, that is
\begin{align}\label{FBDw0}
F^{R}_{BD}(\omega):=\int_{-\infty}^{+\infty}G_{BD}^{+R}(\Delta\tau) e^{-i \omega\Delta\tau} d\Delta\tau=F_{BD}(\omega)-F_{FLA}(\omega)\;.
\end{align}
Here, the  Fourier transform of the Wightman function $G_{BD}^{+}(\Delta\tau)$, which is actually  the response function of a uniformly accelerated particle detector near the reflecting boundary, takes the following form (see Appendix~\ref{appd1} for details)
\begin{equation}\label{F2w0}
F_{BD}(\omega)=\frac{1}{2}\frac{1}{e^{\omega/T_u}+1}\Big[1- P_{\frac{i \omega}{2 \pi T_u}-\frac{1}{2}}\big(8 \pi^{2} T_{u}^{2} L^{2}+1\big)\Big]\;
\end{equation}
with the Unruh temperature $T_{u}=a/2\pi$, and  $P_\nu(x)$ representing  the  associated Legendre function of the first kind~\cite{Gradshteyn:2007}
\begin{equation}\label{PQ}
P_{\nu}(\cosh\alpha):=\frac{\sqrt{2}}{\pi }\int_{0}^{\alpha}\frac{\cosh[(\nu+\frac{1}{2})u]}{(\cosh\alpha-\cosh u)^{\frac{1}{2}}} du , \quad \alpha>0\;,
\end{equation}
while the Fourier transform of Eq.~(\ref{Ffla}) can be straightforwardly carried out as
 \begin{equation}\label{greenw-1}
F_{FLA}(\omega)=\int_{-\infty}^{+\infty}G_{FLA}^{+}(\Delta\tau) e^{-i \omega\Delta\tau} d\Delta\tau=\frac{1}{2}\theta(-\omega)\;
\end{equation}
with $\theta(x)$ representing the Heaviside step function.
Thus, we have
\begin{align}\label{FBDw}
F^{R}_{BD}(\omega)=\frac{1}{2}\frac{1}{e^{\omega/T_u}+1}- \frac{1}{2}\frac{1}{e^{\omega/T_u}+1}P_{\frac{i \omega}{2 \pi T_u}-\frac{1}{2}}\big(8 \pi^{2} T_{u}^{2} L^{2}+1\big)-\frac{\theta(-\omega)}{2}\;.
\end{align}

As we can see that  the first, the second  and the third term in $F^{R}_{BD}(\omega)$ respectively originates from the first, the second term in Eq.~(\ref{Greenpbac}) and the Fourier transform of the Wightman function of the static atom, and therefore
they  respectively  correspond to  terms associated with the accelerated atom, the image of it  and  the static atom.
A substitution of Eq.~(\ref{FBDw}) into Eq.~(\ref{cf-11}) yields the symmetric correlation function for the
field
\begin{align}
C^F_{BD}(x( \tau ),x( \tau'))=-\frac{1}{8\pi}\int_{0}^{\infty} d\omega
\big(e^{{i}\omega{\Delta\tau}}+e^{{-i}\omega{\Delta\tau}}
\big) P_{\frac{i \omega}{2 \pi T_u}-\frac{1}{2}}\big(8 \pi^{2} T_{u}^{2} L^{2}+1\big)\;,
\end{align}
where the identity $P_{\nu}(x)=P_{-\nu-1}(x)$ has been used. Here, the integration of $F_{FLA}(\omega)$  cancels out that over the first  term of $F^{R}_{BD}(\omega)$, i.e., $(2e^{\omega /T_u}+2)^{-1}$ in Eq.~(\ref{FBDw}).
Similarly, the linear susceptibility function of the field works out to
\begin{align}
\chi^F_{BD}(x( \tau ),x( \tau'))=&-\frac{1}{8\pi}\int_{0}^{\infty} d\omega
\Big(e^{{i}\omega{\Delta\tau}}-e^{{-i}\omega{\Delta\tau}}
\Big)\Big\{\tanh \Big(\frac{\omega}{2T_u} \Big)\nonumber\\&\times\Big[1-P_{\frac{i \omega}{2 \pi T_u}-\frac{1}{2}}\big(8 \pi^{2} T_{u}^{2} L^{2}+1\big)\Big]-1\Big\}\;.
\end{align}

Inserting above equations into Eqs.~(\ref{ebvf}) and~(\ref{ebrr}) and extending the range of integration to infinity for sufficiently long time interval, the contributions of vacuum fluctuations and radiation reaction to the energy shift of the level state $|b\rangle$ ($b\in\{-,+\}$) are found to
 be respectively  given by
 \begin{align}\label{pbacvf}
(\delta E_b)_{vf}=&-\frac{\mu^2}{8\pi}\sum_{d}\big|\big\langle b\big|R_{2}(0)\big| d\big\rangle\big|^{2}\int_{0}^{\infty}d\omega{\mathcal{P}}\Big(\frac{1}{\omega+\omega_{bd}}-\frac{1}{\omega-\omega_{bd}}\Big)\nonumber\\&\times P_{\frac{i \omega}{2 \pi T_u}-\frac{1}{2}}\big(8 \pi^{2} T_{u}^{2} L^{2}+1\big)\;,
\end{align}
\begin{align}\label{pbacrr}
(\delta E_b)_{rr}=&-\frac{\mu^2}{8\pi}\sum_{d}\big|\big\langle
b\big|R_{2}(0)\big| d\big\rangle\big|^{2}\int_{0}^{\infty}d\omega
{\mathcal{P}}\Big(\frac{1}{\omega+\omega_{bd}}+\frac{1}{\omega-\omega_{bd}}\Big)\nonumber\\&\times
\Big\{\tanh\Big(\frac{\omega}{2T_u} \Big)\Big[1- P_{\frac{i \omega}{2 \pi T_u}-\frac{1}{2}}\big(8 \pi^{2} T_{u}^{2} L^{2}+1\big)\Big]-1\Big\}\;,
\end{align}
where  $\mathcal{P}$ denotes the principal value integral.
It follows from Eqs.~(\ref{pbacvf}) and~(\ref{pbacrr})  that $(\delta E_+)_{vf}=-(\delta E_-)_{vf}$ and $(\delta E_+)_{rr}=(\delta E_-)_{rr}$. So,  the Lamb shift for the accelerated atom near the reflecting boundary is determined by $(\delta E_+)_{vf}$ only, and  it is given,  after considering $\sum\limits_{d}|\langle b|R_{2}(0)| d\rangle|^{2}=1/4$,  by
\begin{align}\label{deltaBD0}
\Delta_{BD}=-\frac{\mu^2}{16\pi}\int_{0}^{\infty}d\omega{\mathcal{P}}\Big(\frac{1}{\omega+\omega_{0}}-\frac{1}{\omega-\omega_{0}}\Big) P_{\frac{i \omega}{2 \pi T_u}-\frac{1}{2}}\big(8 \pi^{2} T_{u}^{2} L^{2}+1\big)\;.
\end{align}

It is interesting to note that  here only the contribution from  the image of accelerated atom retains, and the contribution from the accelerated atom cancels that of the static atom, meaning that  uniform acceleration does not induce corrections to the Lamb shift in free space without boundary.
Furthermore, from the following properties of the  associated Legendre function of the first kind,
\begin{equation}\label{id-1}
\lim_{x\rightarrow\infty} P_{iy-1/2}(x^2+1)=0,~~y\in\rm{Reals}\;
\end{equation}
and
\begin{equation}\label{id-2}
\lim_{y\rightarrow\infty} P_{iy-1/2}(x^2+1)=0, ~~x \neq 0\;,
\end{equation}
we can see that $\Delta_{BD}$  vanishes when $L\rightarrow\infty$ which means there is no the correction to the Lamb shift of the accelerated atom in a free space in (2+1)-dimensional spacetime. So does the result in the limit of  $a\rightarrow\infty$ (or $T_u\rightarrow\infty$) for a finite distance $L$, meaning that the corrections to the Lamb shift  near a reflecting boundary approach zero as acceleration grows extremely large.


\section{The Lamb shift for a static atom in the BTZ spacetime}
\label{sec4}

In this section, we consider the Lamb shift for a static two-level atom coupled to conformal massless scalar fields in vacuum in the BTZ spacetime of which the line element can be written in Schwarzschild-like coordinates~\cite{Lifschytz:1994,Carlip:1995} as
\begin{equation}
d s^{2}=-\frac{r^2-M\ell^2}{\ell^2} d t^{2}+\frac{\ell^2} {r^2-M\ell^2}d r^{2}+r^{2} d \phi^{2}\;.
\end{equation}
The metric describes an asymptotically anti-de Sitter space with  $\Lambda=-1/\ell^{2}$ a negative cosmological constant, which has a horizon at $r_+ =\sqrt{M}\ell$ with $M$ representing the  mass of the BTZ solution.

To obtain  the Lamb shift in the BTZ spacetime,  we need  the Wightman function of the conformal scalar field  to evaluate  the contributions of vacuum fluctuations and radiation reaction to the atomic energy level shifts.  In this paper, we assume that the scalar field is in  the Hartle-Hawking vacuum in the BTZ spacetime. Note that since the BTZ spacetime can  be obtained by a topological identification of  $AdS_3$ spacetime,  the Wightman function for the conformal massless scalar field in the Hartle-Hawking vacuum in the BTZ spacetime can be expressed in terms of the  corresponding Wightman function in $AdS_3$ spacetime by using the method of images~\cite{Lifschytz:1994,Carlip:1995}
\begin{equation}\label{adssum}
G^{+}_{BTZ}(x,x')=\sum_{n=-\infty}^{+\infty}G^{+}_{AdS}(x,\Gamma^n {x}')\;,
\end{equation}
where $G^{+}_{AdS}(x, {x}')$ is the  Wightman function in $AdS_3$ spacetime and  $\Gamma{x}'$  represents the action of the identification $\phi\mapsto\phi+2\pi $ on  point $x'$.
Assuming, for simplicity, that the field  satisfies the Dirichlet boundary condition at spatial infinity due to lack of global hyperbolicity of the BTZ spacetime,  the Wightman function can be found analytically as follows~\cite{Lifschytz:1994}
 \begin{equation}\label{GBTZ}
G^{+}_{BTZ}(x,x')=\frac{1}{4  \sqrt{2} \pi \ell} \sum_{n=-\infty}^{\infty}\left[\frac{1}{\sqrt{\sigma_{n}}}-\frac{1}{\sqrt{\sigma_{n}+2}}\right]\;,
\end{equation}
where
\begin{equation}
\sigma_{n}:=\frac{{r} r^{\prime}}{r_+^{2}} \cosh
 \left[\sqrt{M}(\Delta \phi-2 \pi n)\right]-1-\frac{\sqrt{\left(r^{2}-r_+^{2}\right)\left(r^{\prime 2}
 -r_{+}^{2}\right)}}{r_{+}^{2}} \cosh \left[\frac{r_+}{\ell^{2}} \Delta
  t\right]\;,
\end{equation}
with $\Delta\phi=\phi-\phi',~\Delta {t}=t-t'$.

In the following discussions, we suppose that the two-level atom is spatially fixed  at a constant $r$ in the BTZ spacetime such that  $\Delta\phi=0,~\Delta\tau=\tau-\tau'=\sqrt{-g_{00}}\Delta {t}$. Since our interest is the relative Lamb shift in the BTZ spacetime, we shall use  the Wightman function which is obtained by subtracting that in the  Minkowski vacuum in (2+1)-dimensional flat spacetime in the subsequent discussions, i.e.,  $G_{BTZ}^{+R}(\Delta\tau):=G_{BTZ}^{+}(\Delta\tau)-G_{FLA}^{+}(\Delta\tau)$.
Then the Fourier transform for the  Wightman function $G_{BTZ}^{+R}(\Delta\tau)$ reads
\begin{equation}\label{re-FourBTZ}
F^{R}_{BTZ}(\omega):=\int_{-\infty}^{+\infty}G_{BTZ}^{+R}(\Delta\tau) e^{-i \omega\Delta\tau} d\Delta\tau=F_{BTZ}(\omega)-F_{FLA}(\omega)\;.
\end{equation}
Note that  $F_{FLA}(\omega)$ is given by Eq.~(\ref{greenw-1}). Here,  the Fourier transform  of the Wightman function~(\ref{GBTZ}), $F_{BTZ}(\omega)$, along the  trajectory of the static atom is given by~\cite{Lifschytz:1994}, which is the response function of a static detector in the BTZ spacetime as well
  \begin{align}\label{FourBTZ}
F_{BTZ}(\omega)=&\int_{-\infty}^{+\infty}G_{BTZ}^{+}(\Delta\tau) e^{-i \omega\Delta\tau} d\Delta\tau
\nonumber\\
=&\frac{1}{2}\frac{1}{e^{\omega /
T_h}+1}\sum_{n=-\infty}^{\infty}\Big[P_{\frac{i \omega}{2 \pi T_h}-\frac{1}{2}}(\cosh
\alpha_{n}) - P_{\frac{i \omega}{2 \pi T_h}-\frac{1}{2}}(\cosh
\beta_{n})\Big]\;,
\end{align}
where $T_h$ is the local  temperature given by
\begin{equation}\label{thd}
T_h=\frac{r_{+}}{2 \pi \ell \sqrt{r^{2}-r_{+}^{2}}}\;,
\end{equation}
 and the auxiliary functions $\cosh
\alpha_{n}$  and $\cosh
\beta_{n}$  are  defined as
\begin{align}\label{alphbeta}
\cosh
\alpha_{n}:=\frac{r_{+}^{2}}{r^{2}-r_{+}^{2}}\Big[\frac{r^{2}}{r_{+}^{2}}
\cosh \Big(2\pi {n} \sqrt{M}\Big)-1\Big]\;,\nonumber\\
\cosh
\beta_{n}:=\frac{r_{+}^{2}}{r^{2}-r_{+}^{2}}\Big[\frac{r^{2}}{r_{+}^{2}}
\cosh \Big(2\pi {n} \sqrt{M}\Big)+1\Big]\;.
\end{align}
Let us note that the local temperature of the BTZ spacetime  can be rewritten in the following form~\cite{Zhjl:2020}
\begin{equation}
T_h=\frac{\sqrt{a_{BTZ}^2-\ell^{-2}}}{2\pi}\;
\end{equation}
with $a_{BTZ}:=r/(\ell\sqrt{r^2-r_+^2})$ representing the  acceleration of the constant $r$ trajectory in the BTZ spacetime. Here $T_h$  is analogous  to the temperature felt by an accelerated observer in $AdS_3$ spacetime~\cite{Jennings:2010}.   It is  demonstrated in Ref.~\cite{Jennings:2010} that  there exists a critical acceleration, $1/\ell$, in $AdS_3$ spacetime, and only the observer with a constant super-critical acceleration $a$ (i.e., $a>1/\ell$) can register the quasi-thermal response with a temperature equal to $\sqrt{a^2-\ell^{-2}}/(2\pi)$. Noteworthily,  the response function~(\ref{FourBTZ})  displays  the Fermi-Dirac distribution as was  noted in Ref.~\cite{Lifschytz:1994}.

Consequently,  we can reexpress the Fourier transform of the Wightman function $G^{+R}_{BD}(x(\tau),x(\tau'))$  as
\begin{align}\label{Fourbtz-r}
F^{R}_{BTZ}(\omega)= F^R_{n=0}(\omega)+F^R_{n\neq0}(\omega)
\end{align}
with
\begin{equation}\label{Fourbtz-r0}
F^R_{n=0}(\omega)=\frac{1}{2}\frac{1}{e^{\omega /
T_h}+1}-\frac{1}{2}\frac{1}{e^{\omega /
T_h}+1}P_{\frac{i \omega}{2 \pi T_h}-\frac{1}{2}}(\cosh\beta_0)-\frac{\theta(-\omega)}{2}\;,
\end{equation}
and
\begin{equation}\label{Fourbtz-rn}
F^R_{n\neq0}(\omega)=\frac{1}{e^{\omega /
T_h}+1}\sum_{n=1}^{\infty}\Big[P_{\frac{i \omega}{2 \pi T_h}-\frac{1}{2}}(\cosh
\alpha_{n}) -P_{\frac{i \omega}{2 \pi T_h}-\frac{1}{2}}(\cosh
\beta_{n})\Big]\;,
\end{equation}
where we have used the identity $P_\nu(1)=1$ for the associated Legendre function
of the first kind and split  $F^R_{BTZ}(\omega)$ into  $n=0$ and $n\neq0$  terms  for convenience.

{With the symmetric correlation function defined in~(\ref{cf-1}), we  have}
\begin{align}\label{CFBTZ}
C^{F}_{BTZ}(x(\tau),x(\tau'))
=&\frac{1}{4\pi} \sum_{n=1}^{\infty}\int_{0}^{\infty} d\omega
\Big(e^{{i}\omega{\Delta\tau}}+e^{{-i}\omega{\Delta\tau}}
\Big)\Big[P_{\frac{i \omega}{2 \pi T_h}-\frac{1}{2}}\left(\cosh
\alpha_{n}\right) - P_{\frac{i \omega}{2 \pi
T_h}-\frac{1}{2}}(\cosh \beta_{n})\Big]\nonumber\\&-\frac{1}{8\pi}\int_{0}^{\infty} d\omega
\Big(e^{{i}\omega{\Delta\tau}}+e^{{-i}\omega{\Delta\tau}}
\Big)P_{\frac{i \omega}{2 \pi
T_h}-\frac{1}{2}}(\cosh \beta_{0})\;.
\end{align}
 {One can see from Eq.~(\ref{CFBTZ}) that the inverse-Fourier integration over $F_{FLA}(\omega)$ also cancels out that over the first  term in $F^R_{n=0}(\omega)$, i.e., $(2e^{\omega /T_h}+2)^{-1}$ in Eq.~(\ref{Fourbtz-r0}).}
Similarly, the linear susceptibility function takes the following form
\begin{align}\label{chifBTZ}
\chi^{F}_{BTZ}(x(\tau),x(\tau'))
=&-\frac{1}{4\pi}\sum_{n=1}^{\infty} \int_{0}^{\infty}d\omega
\Big(e^{{i}\omega{\Delta\tau}}-e^{{-i}\omega{\Delta\tau}} \Big)
\tanh\Big(\frac{\omega}{2T_h}\Big)\Big[P_{\frac{i \omega}{2 \pi{T_h}}-\frac{1}{2}}(\cosh \alpha_{n})\nonumber\\
& - P_{\frac{i \omega}{2 \pi T_h}-\frac{1}{2}}(\cosh
\beta_{n})\Big]-\frac{1}{8\pi}\int_{0}^{\infty}d\omega
\Big(e^{{i}\omega{\Delta\tau}}-e^{{-i}\omega{\Delta\tau}}\Big)
\Big\{\tanh\Big(\frac{\omega}{2T_h}\Big)\nonumber\\
&\times \Big[1- P_{\frac{i \omega}{2 \pi T_h}-\frac{1}{2}}(\cosh
\beta_{0})\Big]-1 \Big\}\;.
\end{align}
Inserting above equations into Eqs.~(\ref{ebvf}) and~(\ref{ebrr}), the contributions of vacuum fluctuations and radiation reaction to the energy shift of the level state $|b\rangle$ are found to be respectively  given by
\begin{align}\label{ebvf-4}
(\delta E_b)_{vf}
=&\frac{\mu^2}{4\pi}\sum\limits_{d}|\langle b|R_{2}(0)| d\rangle|^{2} \sum_{n=1}^{\infty}\int_{0}^{\infty}d\omega
{\mathcal{P}}\Big(\frac{1}{\omega+\omega_{bd}}-\frac{1}{\omega-\omega_{bd}}\Big)\Big[P_{\frac{i
\omega}{2 \pi{T_h}}-\frac{1}{2}}(\cosh \alpha_{n})\nonumber\\
&- P_{\frac{i \omega}{2 \pi {T_h}}-\frac{1}{2}}(\cosh \beta_{n})\Big]-\frac{\mu^2}{8\pi} \sum\limits_{d}|\langle b|R_{2}(0)| d\rangle|^{2} \int_{0}^{\infty}d\omega
{\mathcal{P}}\Big(\frac{1}{\omega+\omega_{bd}}-\frac{1}{\omega-\omega_{bd}}\Big)\nonumber\\
&\times P_{\frac{i \omega}{2 \pi {T_h}}-\frac{1}{2}}(\cosh
\beta_{0})\;,
\end{align}
and
\begin{align}\label{ebrr-4}
(\delta E_b)_{rr}=&-\frac{\mu^2}{4\pi}\sum\limits_{d}|\langle b|R_{2}(0)| d\rangle|^{2}\sum_{n=1}^{\infty}\int_{0}^{\infty}d\omega
{\mathcal{P}}\Big(\frac{1}{\omega+\omega_{bd}}+\frac{1}{\omega-\omega_{bd}}\Big)\tanh\Big(\frac{\omega}{2T_h}\Big)\nonumber\\
&\times \Big[P_{\frac{i \omega}{2 \pi {T_h}}-\frac{1}{2}}(\cosh \alpha_{n})
-P_{\frac{i \omega}{2 \pi {T_h}}-\frac{1}{2}}(\cosh
\beta_{n})\Big]-\frac{\mu^2}{8\pi}\sum\limits_{d}|\langle b|R_{2}(0)| d\rangle|^{2}\nonumber\\
&\times\int_{0}^{\infty}d\omega
{\mathcal{P}}\Big(\frac{1}{\omega+\omega_{bd}}+\frac{1}{\omega-\omega_{bd}}\Big)\Big{\{} \tanh\Big(\frac{\omega}{2T_h}\Big)\Big[1-P_{\frac{i \omega}{2 \pi {T_h}}-\frac{1}{2}}(\cosh
\beta_{0})\Big]-1\Big{\}}\;.
\end{align}
So, it is easy to get that the Lamb shift for the two-level atom  in the BTZ spacetime is determined by $(\delta E_+)_{vf}$ only,  that is
\begin{align}\label{Delta-BTZall}
\Delta_{BTZ}=2(\delta E_+)_{vf}=&\Delta_{n=0}+\Delta_{n\neq0}\;
\end{align}
with
\begin{align}\label{beta0}
\Delta_{n=0}:=&-\frac{\mu^2}{16\pi}\int_{0}^{\infty}d\omega
{\mathcal{P}}\Big(\frac{1}{\omega+\omega_0}-\frac{1}{\omega-\omega_0}\Big)
P_{\frac{i \omega}{2 \pi T_h}-\frac{1}{2}}(\cosh \beta_{0})\nonumber\\
=&-
\frac{\mu^2}{16\pi}\int_{0}^{\infty}d\omega
{\mathcal{P}}\Big(\frac{1}{\omega+\omega_0}-\frac{1}{\omega-\omega_0}\Big)
P_{\frac{i \omega}{2 \pi T_h}-\frac{1}{2}}\big(8 \pi^{2} T_{h}^{2} \ell^{2}+1\big)\;,
\end{align}
and
\begin{align}\label{be-alh-n}
\Delta_{n\neq 0}:=\frac{\mu^2}{8\pi}\sum_{n=1}^{\infty} \int_{0}^{\infty}d\omega
{\mathcal{P}}\Big(\frac{1}{\omega+\omega_0}-\frac{1}{\omega-\omega_0}\Big)\Big[P_{\frac{i
\omega}{2 \pi T_h}-\frac{1}{2}}(\cosh \alpha_{n})  - P_{\frac{i
\omega}{2 \pi T_h}-\frac{1}{2}}(\cosh \beta_{n})\Big]\;.
\end{align}
Here  $\Delta_{n=0}$ represents the contribution from the second term of   the response function $F^R_{n=0}(\omega)$ in~(\ref{Fourbtz-r}), and $\Delta_{n\neq 0}$ results from  $F^R_{n\neq0}(\omega)$.

{Recalling  the Lamb shift of the accelerated atom near the reflecting boundary, $\Delta_{BD}$ (i.e., Eq.~(\ref{deltaBD0})),
 we find that Eqs.~(\ref{deltaBD0}) and ~(\ref{beta0}) become the same if one replaces $T_u$ and $L$ in Eq.~(\ref{deltaBD0})  with $T_h $ and $\ell$ (or equivalently, let $a \equiv \sqrt{a_{BTZ}^2-\ell^{-2}}$ and $L \equiv \ell$).
  } In fact, this  equivalence between $\Delta_{n=0}$ and $\Delta_{BD}$  can also be seen from the response functions,   since the $n=0$ term in the response function~(\ref{FourBTZ}) in the BTZ spacetime, i.e.,
\begin{equation}\label{Fneq0}
\frac{1}{2}\frac{1}{e^{\omega/T_h}+1}\Big[1- P_{\frac{i \omega}{2 \pi T_h}-\frac{1}{2}}(8 \pi^{2} T_{h}^{2} \ell^{2}+1)\Big]\;,
\end{equation}
coincides with Eq.~(\ref{F2w0}).

Let us now analyze the contribution of $n\neq 0$ terms in the response function to the Lamb shift. Considering that the auxiliary functions in Eq.~(\ref{alphbeta}) can be rewritten  as
\begin{align}\label{alphbeta2}
&\cosh\alpha_{n}=8 \pi^{2} T_{h}^{2}{r^{2}\sinh ^{2} \big(n \pi\sqrt{M}\big)}/{M}+1\;,\nonumber\\
&\cosh\beta_{n}=8 \pi^{2} T_{h}^{2}\Big[{\frac{r^2}{M}\sinh ^{2} \big(n \pi\sqrt{M}\big)}+\ell^{2}\Big]+1\;,
\end{align}
we have
\begin{align}\label{Deltaneq02}
\Delta_{n\neq 0}=&\frac{\mu^2}{8\pi}\sum_{n=1}^{\infty}\;\Bigg[\int_{0}^{\infty}d\omega
{\mathcal{P}}\Big(\frac{1}{\omega+\omega_0}-\frac{1}{\omega-\omega_0}\Big)P_{\frac{i
\omega}{2 \pi T_h}-\frac{1}{2}}(8 \pi^{2} T_{h}^{2} L_{\alpha_n}^{2}+1) \nonumber\\&-\int_{0}^{\infty}d\omega
{\mathcal{P}}\Big(\frac{1}{\omega+\omega_0}-\frac{1}{\omega-\omega_0}\Big)P_{\frac{i
\omega}{2 \pi T_h}-\frac{1}{2}}(8 \pi^{2} T_{h}^{2} L_{\beta_n}^{2}+1)\Bigg]\;
\end{align}
with $L_{\alpha_n}:={r\sinh\big(n \pi \sqrt{M} \big) }/\sqrt{M}$ and $L_{\beta_n}:=[{r^{2}\sinh ^{2} \big(n \pi\sqrt{M}\big)}/{M}+\ell^{2}]^{1/2}$.  A comparison of the above result with Eq.~(\ref{deltaBD0})  suggests that the contribution of each $ {n\neq 0}$ term to the Lamb shift can be understood  as that from a pair of a source, which is accelerated along a reflecting boundary at an effective distance to $L_{\alpha_n}$ in a flat spacetime with an acceleration $ a=2\pi{T_h}=\sqrt{a_{BTZ}^2-\ell^{-2}}=\sqrt{M/(r^2-M\ell^2)}$,  and an image  with the same acceleration  but at a different effective distance $L_{\beta_n}$.

Now, let  us turn our attention to analytically evaluating the Lamb shift $\Delta_{BTZ}$ in some special cases.  First,  for the case of the atom  in the asymptotic region far form the horizon, i.e., $r \gg r_{+}$. In this case,  the temperature $T_h$ is very small while  both $T_h^2L_{\alpha_n}^2$ and  $T_h^2L_{\beta_n}^2$ approach  $\sinh^2(n\pi\sqrt{M})/(4\pi^2)$,  and then according to Eq.~(\ref{id-2}), we  only need to take $\Delta_{n=0}$ into account, which  is now given by (see Appendix \ref{appd2} for details)
\begin{equation}\label{Delta-BTZ-ap10}
\Delta_{BTZ}\approx \Delta_{n=0}\approx -\frac{\mu^2}{16}\Big[H_0(2\omega_0 \ell)-\frac{r_{+}^2}{r^{2}}f(2\omega_0\ell)\Big]\;,
\end{equation}
with
\begin{equation}
f(2\omega_0 \ell)=\frac{1}{12\omega_0 \ell}\Big[4\omega_0^2
\ell^{2}H_{-1}(2\omega_0 \ell)+4\omega_0 \ell
H_0(2\omega_0 \ell)-H_1(2\omega_0 \ell)\Big]\;,
\end{equation}
which can be further expressed, in the limit of $r \rightarrow \infty$ (or $a_{BTZ}\rightarrow 1/\ell$) as
\begin{equation}\label{Delta-BTZ-ap11}
\Delta_{BTZ}=-\frac{\mu^2}{16}H_0(2\omega_0 \ell)\approx\left\{ \begin{aligned}
        &-\frac{ \mu^2 \omega_0\ell}{4\pi}\;,\quad&\omega_0\ell\ll1 \;;\\
        &
        -\frac{\mu^2}{16\sqrt{\pi\omega_0\ell}}\sin\Big(2\omega_0\ell-\frac{\pi}{4}\Big)\;,\quad&\omega_0\ell\gg1\;.
        \end{aligned} \right.
\end{equation}
 It is worthwhile to note that $\Delta_{BTZ}$ in the asymptotic region is negative when  $\omega_0\ell\ll1$ or $\ell/\lambda _0\ll1$. Here $\lambda_0:=1/\omega_0$ denotes the transition wavelength of the atom. This means that the Lamb shift  is always suppressed in the BTZ spacetime, while  it can be either positive or negative  when $\omega_0\ell\gg1$ or $\ell/\lambda _0\gg1$, signaling either enhancement or suppression contingent on the value of $\omega_0\ell$.

When the atom is in the region near the horizon of the BTZ spacetime, i.e., ${r-r_+}\ll{r_+}$, since both $L_{\alpha_{n}}$ and $L_{\beta_{n}}$ are finite and the temperature $T_h$ is extremely large, $\Delta_{n\neq 0}$ is also negligible. So again, only $\Delta_{n=0}$  needs to be considered  for the Lamb shift,
which can now be approximated as  (see Appendix~\ref{appd2} for details)
 \begin{align}\label{DeltaBTZML}
\Delta_{BTZ}\approx -\frac{\mu^{2} \omega_0 \ell (r-r_{+}) }{16\pi r_{+}} \Big[ \ln\Big(\frac{r-r_{+}}{2r_{+}}\Big)\Big]^2\;.
\end{align}
Interestingly, as the atom  approaches the horizon, i.e.,  $r\rightarrow r_+$, the above results suggest that $\Delta_{BTZ}$ would vanish, implying that the Lamb shift reduces to that in the flat spacetime. This is remarkable since the temperature $T_h$  or acceleration $a_{BTZ}$ blows up as the horizon is approached.

\begin{figure}[!htbp]
\centering
\subfloat[$\ell/\lambda_{0}=0.05$]{\label{rd11}\includegraphics[width=0.325\linewidth]{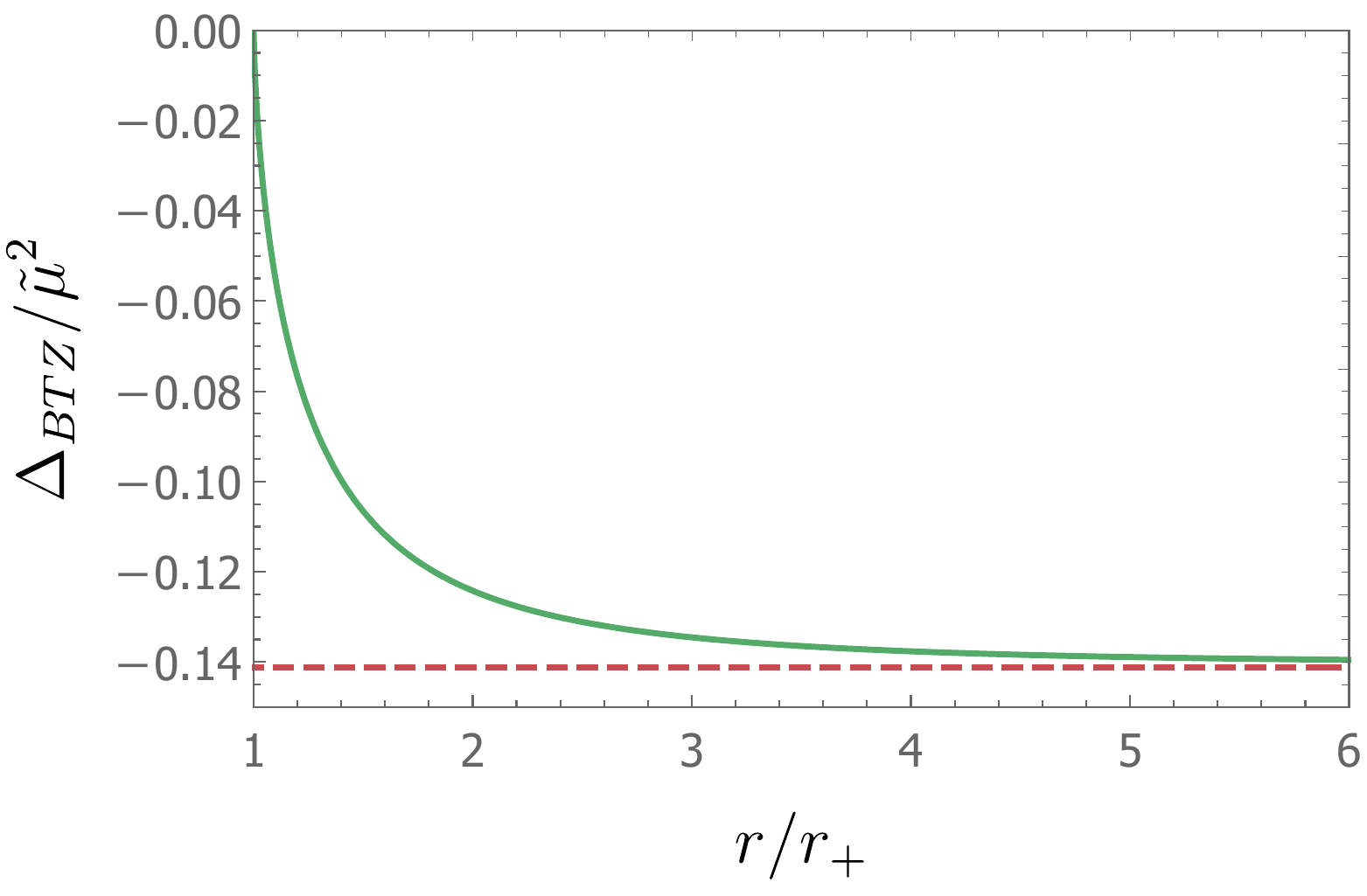}}
\subfloat[$\ell/\lambda_{0}=6.00$]{\label{rd12}\includegraphics[width=0.325\linewidth]{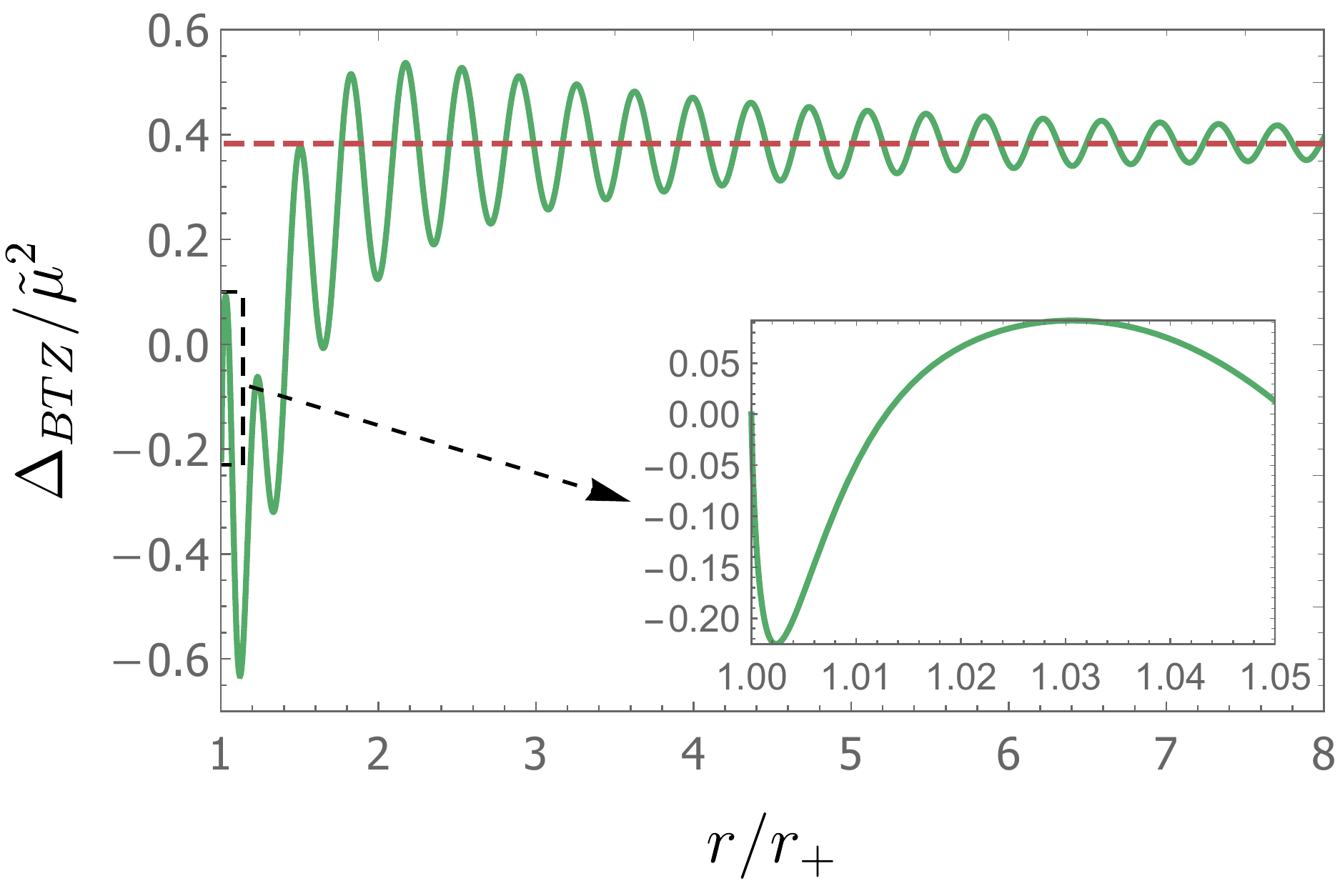}}
\subfloat[$\ell/\lambda_{0}=7.00$]{\label{rd13}\includegraphics[width=0.325\linewidth]{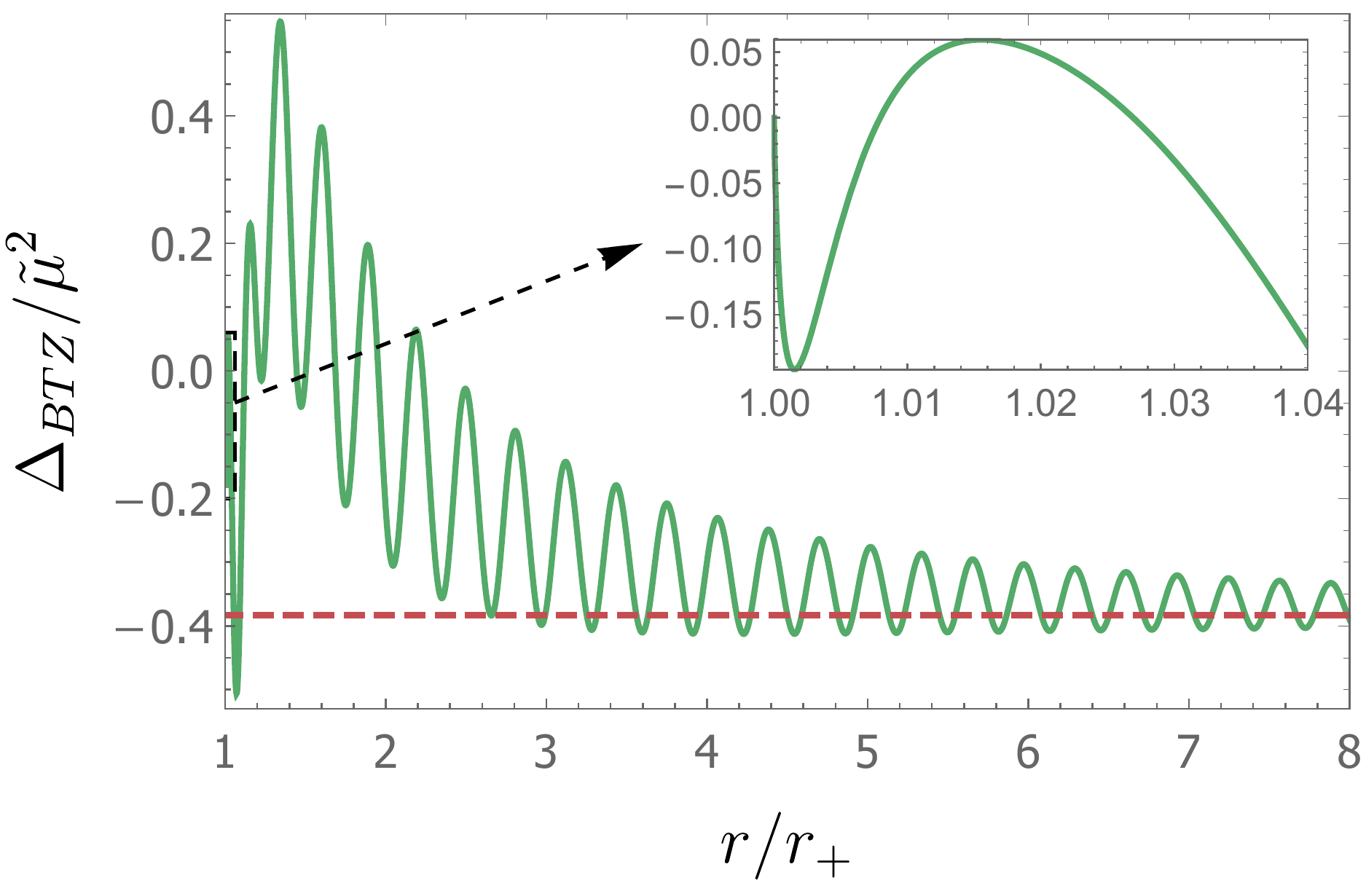}}
\caption{The Lamb shift is plotted as a function of $r/r_{+}$  for (a) $\ell/\lambda_{0}=0.05$, (b) $\ell/\lambda_{0}=6.00$, (c) and $\ell/\lambda_{0}=7.00$ with the transition wavelength of the atom defined by $\lambda_0=1/\omega_0$. Here, we have taken $\tilde{\mu}^2={\mu^2}/{(8\sqrt{2}\pi)}$ and $M=0.20$.  The dashed red lines in all plots indicate the corresponding convergent values of $\Delta_{BTZ}/\tilde{\mu}^2$ (i.e., approximately equal to $ -0.1413$ in (a), $ 0.3833$ in (b), and $-0.3831$ in (c)) in the limit of $r\rightarrow\infty$. }\label{rd1}
\end{figure}

We now examine the properties of the Lamb shift at a generic spatial position $r$.  We will  resort to  numerical calculation  since  exact analytical results are unobtainable. In Fig.~(\ref{rd1}), the  Lamb shift is plotted as a function of  the spatial position  $r/r_{+}$ with  given  mass $M$ and $\ell/\lambda_0$.   As shown in Fig.~(\ref{rd1}), the  Lamb shift
 approaches  a fixed  value, i.e.,
$-\mu^2H_0(2\ell/\lambda_0)/16$, in the asymptotic region of the
BTZ spacetime,  which is always negative for $\ell/\lambda_0\ll1$ (see Fig.~(\ref{rd11})) but can be either positive or negative for $\ell/\lambda_0\gg1$ depending on the value of  $\ell/\lambda_0$ (see Fig.~(\ref{rd12})~(\ref{rd13})).  This is in agreement with  our previous  results obtained by analytical analysis in Eq.~(\ref{Delta-BTZ-ap11}).  Moreover,  at $r/r_+\sim1$, all plots in Fig.~(\ref{rd1}) reveal  a vanishing  Lamb shift, and  this  is again in accordance with our analytical result (Eq.~(\ref{DeltaBTZML})) very close to the horizon.

Notably, Fig.~(\ref{rd11}) also indicates that the Lamb shift is always suppressed in the BTZ spacetime as compared to that in the flat spacetime as long as  $\ell/\lambda_0\ll1$, i.e., when the transition wavelength of the atom is much larger than the $AdS$ radius of the BTZ spacetime,  not just in the asymptotic region as our analytical result may seem to suggest. However, if  the atomic  transition wavelength is much less than the $AdS$ radius, i.e., $\ell/\lambda_0\gg1$, the Lamb shift exhibits obvious  oscillations as the atom is moved away from the horizon.  Moreover,   Fig.~(\ref{rd12}) shows that, for a given value of $\ell/\lambda_0$  that gives a positive asymptotic Lamb shift, the Lamb shift may be significantly suppressed with a few oscillations  in the vicinity of the horizon as compared to that in the flat spacetime and oscillates around the positive asymptotic value as the atom is moved far away from the horizon of the BTZ spacetime, while Fig.~(\ref{rd13}) demonstrates that, for a given value of $\ell/\lambda_0$  that leads to a negative asymptotic Lamb shift, the Lamb shift may  instead be significantly enhanced with a few oscillations   in the vicinity of the horizon as compared to that in the flat spacetime and oscillates around the negative asymptotic value. Finally, let us note that the qualitative behaviors of the Lamb shift are not very sensitive of the value of $M$ which is  dimensionless in the BTZ spacetime. So, a value of 0.2 is chosen in our plots of numerical computations.


\section{Conclusion}
\label{sec5}

We have studied  the Lamb shift of a two-level atom arising from its
coupling to the conformal massless scalar field, which satisfies the Dirichlet boundary conditions,  in the Hartle-Hawking vacuum in the BTZ spacetime, and compared it with that of the uniformly accelerated atom near a  perfectly reflecting boundary in the (2+1)-dimensional flat spacetime.

We demonstrate that,  for a spatially fixed atom in the BTZ spacetime, its Lamb shift is structurally similar to that of a uniformly accelerated atom near a  perfectly reflecting boundary in the (2+1)-dimensional  flat spacetime with an acceleration $a=2\pi T_h$ where $T_h$ is the local temperature of the BTZ spacetime. Both analytical analysis and numerical computation show that  the Lamb shift is suppressed  in the BTZ spacetime as compared to that  in the flat spacetime as long as the transition wavelength of the atom is much larger than the $AdS$ radius while it can be either suppressed or enhanced if  the transition wavelength of the atom is much less than the $AdS$ radius. In contrast, the Lamb shift is always suppressed very close to the horizon of the BTZ spacetime and remarkably it rolls back to that in the flat spacetime as the horizon is approached although the local temperature diverges there.

\begin{acknowledgments}
 We would like to thank Jiawei Hu for helpful discussions. This work was supported in part by the NSFC under Grants  No.12075084 and No.12175062,
 the Research Foundation of Education Bureau of Hunan Province, China under Grant
 No.20B371.
\end{acknowledgments}

\appendix
\section{The derivation of  $F_{BD}(\omega)$ }\label{appd1}

According to the definition of the Fourier transform,  we have
\begin{align}\label{F2w}
	F_{BD}(\omega)=&\int_{-\infty}^{+\infty}G_{BD}^{+}(\Delta\tau) e^{-i \omega\Delta\tau} d\Delta\tau\nonumber\\
	=&\frac{a}{8\pi}\int_{-\infty}^{+\infty} \frac{e^{-i \omega \Delta\tau} d\Delta\tau}{\sqrt{-\big[\sinh\big(\frac{a \Delta\tau}{2}-i\epsilon \big) \big]^{2}}}-\frac{\sqrt{2} a}{8\pi}\int_{-\infty}^{+\infty} \frac{e^{-i \omega \Delta\tau} d\Delta\tau}{\sqrt{\cosh\beta_{B}-\cosh(a \Delta\tau-i \epsilon)}}\;,
\end{align}
with $\cosh{\beta_{B}}:=2a^{2}L^{2}+1$.  The first integral  in  Eq.~(\ref{F2w}) can be obtained as follows:
\begin{align}\label{FBD11}
	&\frac{a}{8\pi} \int_{-\infty}^{+\infty} \frac{e^{-i \omega \Delta\tau} d\Delta\tau}{\sqrt{-\big[\sinh\big(\frac{a \Delta\tau}{2}-i\epsilon \big) \big]^{2}}}\nonumber\\
	&=\frac{i a}{8\pi} \int_{-\infty}^{0} e^{-i \omega \Delta\tau}  \Big| \sinh \Big(\frac{a  \Delta\tau}{2}-i\epsilon \Big) \Big|^{-1} d\Delta\tau
	- \frac{i a}{8\pi}\int_{0}^{\infty} e^{-i \omega \Delta\tau}  \Big| \sinh \Big(\frac{a  \Delta\tau}{2}-i\epsilon \Big) \Big|^{-1}d \Delta\tau\nonumber\\
	&=\frac{a}{8\pi}\int_{-\infty}^{\infty} e^{-i \omega  \Delta\tau}\Big[ i \sinh \Big(\frac{a  \Delta\tau}{2}\Big) +\epsilon  \Big]^{-1}d \Delta\tau\nonumber\\
	&=\frac{1}{2}\frac{1}{e^{2\pi \omega/a}+1}\;,
\end{align}
where in the second line an appropriate branch of the square root cut along the negative real axis has been considered via $i \epsilon$ factor (see similar treatments in Ref.~\cite{Lifschytz:1994}).
Similarly, the second integral in Eq.~(\ref{F2w}) can be performed
\begin{align}\label{FBD12}
	&-\frac{\sqrt{2} a}{8\pi}\int_{-\infty}^{+\infty} \frac{e^{-i \omega \Delta\tau} d\Delta\tau}{\sqrt{\cosh\beta_{B}-\cosh(a \Delta\tau-i \epsilon)}}\nonumber\\
	&=\frac{\sqrt{2}}{8\pi i }\int_{-\infty}^{-\beta_{B}} e^{-\frac{i \omega u}{a}} (\cosh u-\cosh\beta_{B})^{-\frac{1}{2}}du
	-\frac{\sqrt{2}}{ 8\pi }\int_{-\beta_{B}}^{\beta_{B}} e^{-\frac{i \omega u}{a}} (\cosh \beta_{B}-\cosh u)^{-\frac{1}{2}}du\nonumber\\
	&\quad-\frac{\sqrt{2}}{ 8\pi i }\int_{\beta_{B}}^{\infty} e^{-\frac{i \omega u}{a}} (\cosh u-\cosh\beta_{B})^{-\frac{1}{2}}du\nonumber\\
	&=-\frac{i }{4\pi} Q_{-\frac{i \omega}{a}-\frac{1}{2}}(\cosh\beta_B)-\frac{1}{4} P_{\frac{i \omega}{a}-\frac{1}{2}}(\cosh\beta_B)+\frac{i }{4 \pi} Q_{\frac{i \omega}{a}-\frac{1}{2}}(\cosh\beta_B)\nonumber\\
	&=-\frac{1}{2}\frac{1}{e^{2\pi \omega/a}+1}{P}_{\frac{i \omega}{a}-\frac{1}{2} } (\cosh\beta_{B})\;,
\end{align}
where $Q_\nu(x)$ denotes the  second associated Legendre function and the identity, $Q_\nu(x)-Q_{-\nu-1}(x)=\pi\cot(\nu\pi)P_{\nu}(x)$,  has been used in the last line.
Combining  Eqs.~(\ref{FBD11}) and ~(\ref{FBD12}), we have
\begin{align}
	F_{BD}(\omega)=&\frac{1}{2}\frac{1}{e^{2\pi \omega/a}+1}\Big[1-{P}_{\frac{i \omega}{a}-\frac{1}{2} }(2 a^{2} L^{2}+1)\Big]\nonumber\\
=&\frac{1}{2}\frac{1}{e^{\omega/T_u}+1}\Big[1- P_{\frac{i \omega}{2 \pi T_u}-\frac{1}{2}}\big(8 \pi^{2} T_{u}^{2} L^{2}+1\big)\Big]\;,
\end{align}
where $T_u=a/(2\pi)$.

\section{Approximations of $\Delta_{BTZ}$}\label{appd2}
Following the definition of the  associated Legendre function of the first kind~(\ref{PQ}), Eq.~(\ref{beta0}) can be rewritten as
\begin{align}\label{beta000}
\Delta_{n=0}&=-\frac{\sqrt{2}\mu^2}{16\pi^2}\int_{0}^{\beta_{0}}du \int_{0}^{\infty}d\omega {\mathcal{P}}\Big(\frac{1}{\omega+\omega_0}-\frac{1}{\omega-\omega_0}\Big) \frac{\cosh\big(\frac{i \omega u}{2 \pi T_h }\big)}{(\cosh\beta_{0}-\cosh u)^{\frac{1}{2}}}\nonumber\\
&=-\frac{\sqrt{2}\mu^2}{16\pi^2}\int_{0}^{\beta_{0}}du \int_{- \infty}^{\infty }d\omega {\mathcal{P}}\Big(\frac{1}{\omega+\omega_0}\Big) \frac{\cos\big(\frac{\omega u}{2 \pi T_h}\big)}{(\cosh\beta_{0}-\cosh u)^{\frac{1}{2}}}\nonumber\\
&=-\frac{\sqrt{2} \mu^2 \beta_{0}}{16\pi}\int_{0}^{1}ds
\frac{\sin\big(\frac{ \omega_0 s \beta_{0}}{2 \pi T_h}\big)
}{\sqrt{\cosh\beta_{0}-\cosh (s\beta_{0})}}\;,
\end{align}
where we used
$$
\int_{-\infty}^{\infty}d\omega  \cos\Big(\frac{\omega u}{2 \pi T_h}\Big) {\mathcal{P}}\Big(\frac{1}{\omega+\omega_0}\Big)=\pi\sin\Big(\frac{\omega_0u}{2 \pi T_h}\Big)\nonumber
$$
and $s=u/\beta_0$ in the last line.

For the case of $r\gg{r_+}$, after considering $T_{h}=\sinh(\beta_{0}/2)/(2\pi \ell)$ and  a power series for the integrand in Eq.~(\ref{beta000})
about small $\beta_0$, we have
\begin{align}\label{DeltaBTZbeta0}
\Delta_{BTZ}\approx\Delta_{n=0}=
&-\frac{\sqrt{2} \mu^2 \beta_{0}}{16\pi}\int_{0}^{1}ds
\frac{\sin\Big[ \frac{\omega_{0} \ell s \beta_{0}}{\sinh(\beta_{0}/2)}\Big] }{\sqrt{\cosh\beta_{0}-\cosh (s\beta_{0})}}\nonumber\\
\approx & -\frac{\mu^2}{8 \pi} \int_{0}^{1}ds \frac{\sin(2\omega_{0}\ell s)}{\sqrt{1-s^{2}}}
+\frac{\mu^{2} \omega_{0} \ell \beta_{0}^{2}}{96\pi}\int_{0}^{1}ds  \frac{s\cos(2\omega_{0}\ell s)}{\sqrt{1-s^{2}}}\nonumber\\
&+\frac{\mu^2 \beta_{0}^{2}}{192 \pi} \int_{0}^{1}ds  \frac{(1+s^{2})\sin(2\omega_{0}\ell s)}{\sqrt{1-s^{2}}}\nonumber\\
\approx & -\frac{\mu^2}{16}\Big[H_0(2\omega_0 \ell)- \frac{r_{+}^{2}}{r^{2}} f(2\omega_0 \ell)\Big]\;,
\end{align}
where
\begin{equation}
f(2\omega_0 \ell)=\frac{1}{12\omega_0 \ell}\Big[4\omega_0^2
\ell^{2}H_{-1}(2\omega_0 \ell)+4\omega_0 \ell
H_0(2\omega_0 \ell)-H_1(2\omega_0 \ell)\Big]\;,
\end{equation}
with $H_{n}(x)$ denoting the Struve function.

For the case of $({r-r_+})\ll{r_+}$,   $\beta_0$  will be extremely large.
Thus, we  have
\begin{align}\label{betaeq}
\Delta_{n=0}=&-\frac{\sqrt{2} \mu^2 \beta_{0}}{16\pi}\int_0^{1}ds\frac{\sin\big(\frac{\omega_{0} s\beta_0}{2
\pi T_h}\big) }{\sqrt{\cosh\beta_0-\cosh (s\beta_0)}}\nonumber\\
\approx&-\frac{\mu^2 \beta_{0}}{16\pi}\int_0^{1}ds\frac{\sin\big[\omega_{0} s\beta_0/(2
\pi T_h)\big] }{\sinh(\beta_0/2)}\nonumber\\
= &-\frac{\mu^2T_h}{8\omega_0\sinh(\beta_0/2)}\Big[1-\cos\Big(\frac{\omega_{0}\beta_0}{2
\pi T_h}\Big)\Big]\;.
\end{align}
According to Eq.~(\ref{betaeq}), the Lamb shift can be further written as
\begin{equation}
\Delta_{BTZ}\approx\Delta_{n=0}\approx-\frac{\mu^{2} \omega_0 \ell (r-r_{+}) }{16\pi r_{+}} \Big[ \ln\Big(\frac{r-r_{+}}{2r_{+}}\Big)\Big]^2\;.
\end{equation}



\end{document}